\documentclass[10pt,conference]{IEEEtran}
\IEEEoverridecommandlockouts
\ifCLASSINFOpdf
\else
\fi
\usepackage{lipsum}
\usepackage{textcomp}
\usepackage{graphicx}
\usepackage{float}
\usepackage{cite}
\usepackage{booktabs}
\usepackage{epstopdf}
\usepackage{multirow}
\usepackage{boldline}
\usepackage{amsmath}
\usepackage{relsize}
\usepackage{array}
\usepackage{colortbl}
\usepackage{algorithm}
\usepackage{algpseudocode}
\usepackage{placeins}
\usepackage{tabulary}
\usepackage{makecell}
\usepackage{enumitem}
\usepackage{mathtools}
\usepackage{soul}
\usepackage{subcaption}
\usepackage{parskip}
\usepackage[normalem]{ulem}

\newcommand\soo{\bgroup\markoverwith{\textcolor{red}{\rule[0.5ex]{2pt}{0.4pt}}}\ULon}

\makeatletter
\newlength \figwidth
\if@twocolumn
\else
\fi
\makeatother

\makeatletter
\def\ps@IEEEtitlepagestyle{%
  \def\@oddfoot{\mycopyrightnotice}%
  \def\@evenfoot{}%
}
\def\mycopyrightnotice{%
  {\footnotesize \textit{This extended abstract was presented in the IEEE International Conference on Communications (ICC) Workshop} $\mathit{2018}$\hfill}
  \gdef\mycopyrightnotice{}
}
\makeatother
\begin{document}
\IEEEoverridecommandlockouts
%

\title{Towards Smart Wireless Body-Centric Networks}

\author{\IEEEauthorblockN{Samiya~M.~Shimly}
\IEEEauthorblockA{The Australian National University \& CSIRO Data61\\
Email: Samiya.Shimly@data61.csiro.au}\vspace{-20pt}
\and
\IEEEauthorblockN{David~B.~Smith}
\IEEEauthorblockA{CSIRO Data61 \& The Australian National University\\
Email: David.Smith@data61.csiro.au}\vspace{-20pt}}

%

\maketitle
\begin{abstract}
We investigate the existence of `long-memory' or long-range dependence (LRD) of the wireless body-centric channels, e.g., on-body, body-to-body (B2B), with real-life experimental dataset collected from 10 co-located wireless body area networks or BANs (people fitted with wearable sensors). We examine two different factors on that purpose such as: the pattern of the decaying autocorrelation function (ACF) and the Hurst exponent. From the experimental outcome, we show that, the ACF decay of the body-centric channels follows a power-like decay and the channels have a Hurst exponent much greater than 0.5 on average. These results indicate that the body-centric channels can possess long-memory or LRD characteristic which can be used for predictive analysis and intelligent decision making to build futuristic wireless human-centered networks that can sense and act autonomously. We also clarify whether the presence of the LRD property is sufficient for reliable prediction of the body-centric channels. 
\end{abstract}
%
\IEEEpeerreviewmaketitle

\vspace{-1em}\section{Introduction}
\vspace{-1em}Wireless body-centric communications are attracting a lot of attention due to the low-cost, suitable new technology for establishing human-to-human or body-to-body networks (BBNs) through wearable sensors. BBNs are envisioned to be self-organizing, smart, and mobile networks that can create their own centralized/decentralized network connection without any external coordination for serving different medical and non-medical applications \cite{sshimly2018}. This type of autonomous decision making activity requires systematic prediction and modeling of the channel behavior which further depends on the `long-memory' characteristic of the channel. Here, we aim to address the following issues:

\begin{itemize}
\setlength\itemsep{-0.5em}
\item{\textit{What is `long-memory' and why is it important?}}
\item{\textit{Do wireless body-centric channels have long-memory?}}
\item{\textit{Is having long-memory sufficient for making reliable prediction?}}
\end{itemize}

\section{Experimental Scenario}
\begin{figure}[!h]
\centering{\includegraphics[width=40mm]{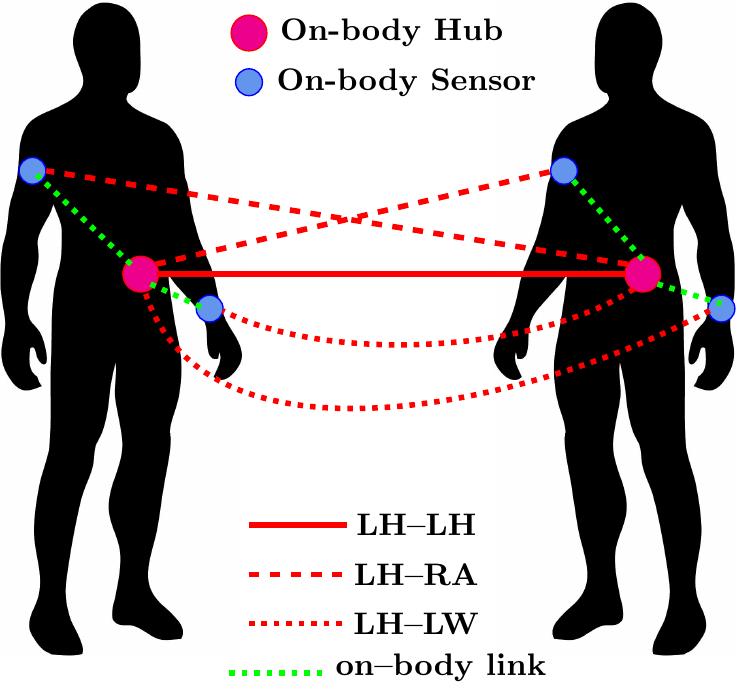}}
\caption{Different body-centric links (on-body, B$2$B) between two BANs wearing on-body hub at the left hip (LH) and two on-body sensors at the right upper arm (RA) and left wrist (LW), respectively.}
\label{b2b}
\vspace{-1.5em}
\end{figure}

\vspace{-1em}We use an open-access dataset which consists of contiguous extensive intra-BAN (on-body) and inter-BAN (body-to-body) channel gain data of around $45$ minutes, captured from $10$ closely located mobile subjects (adult male and female) with a sampling rate of $20$ Hz. 
Each subject wore $1$ transmitter (Tx hub) on the left-hip and $2$ receivers (sensors/ relays) on the left-wrist and  right-upper-arm, respectively (Fig. \ref{b2b}). A description of these wearable radios can be found in \cite{hanlen2010open} and the ``open-access" dataset can be downloaded from \cite{smith2012body}.

\section{Long-memory or Long-range dependence}
Long-memory or Long-range dependence (LRD) is the level of statistical dependence between two points in the time series. The `memory' refers to how strongly the past can influence the future or, how useful is the past data to predict the future consequences. If a channel possesses long-range-dependence then it is more predictable as more data can be used to predict the future.

\subsection{Decaying ACF} 

\vspace{-0.5em}A rough analysis of the dependence is to examine the pattern of the decaying autocorrelation function (ACF) of the channel. For a short-memory process, the dependence between two points decreases rapidly with the increase in time difference, hence the ACF has an exponential decay (faster decay) or drops to $0$ after a certain time lag. On the other hand, if the channel possesses long-memory the ACF decays more slowly (power-like) than an exponential decay.
\begin{figure*}[!htb]
\begin{subfigure}{.3\textwidth}
\centering
\includegraphics[width=.8\textwidth]{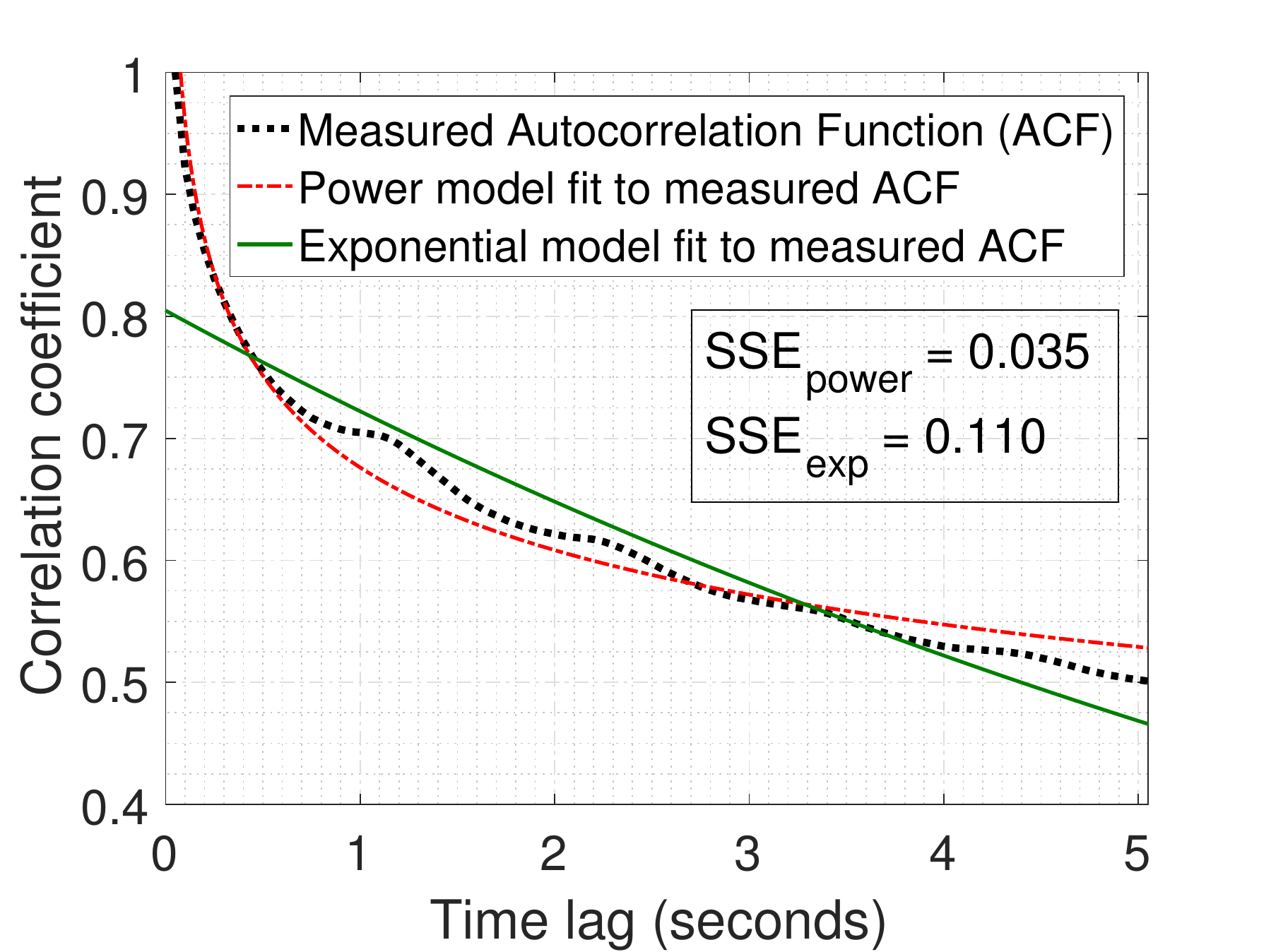}
\caption{ACF (LH--LH)}
\end{subfigure}\hfill
\begin{subfigure}{.3\textwidth}
\centering
\includegraphics[width=.8\textwidth]{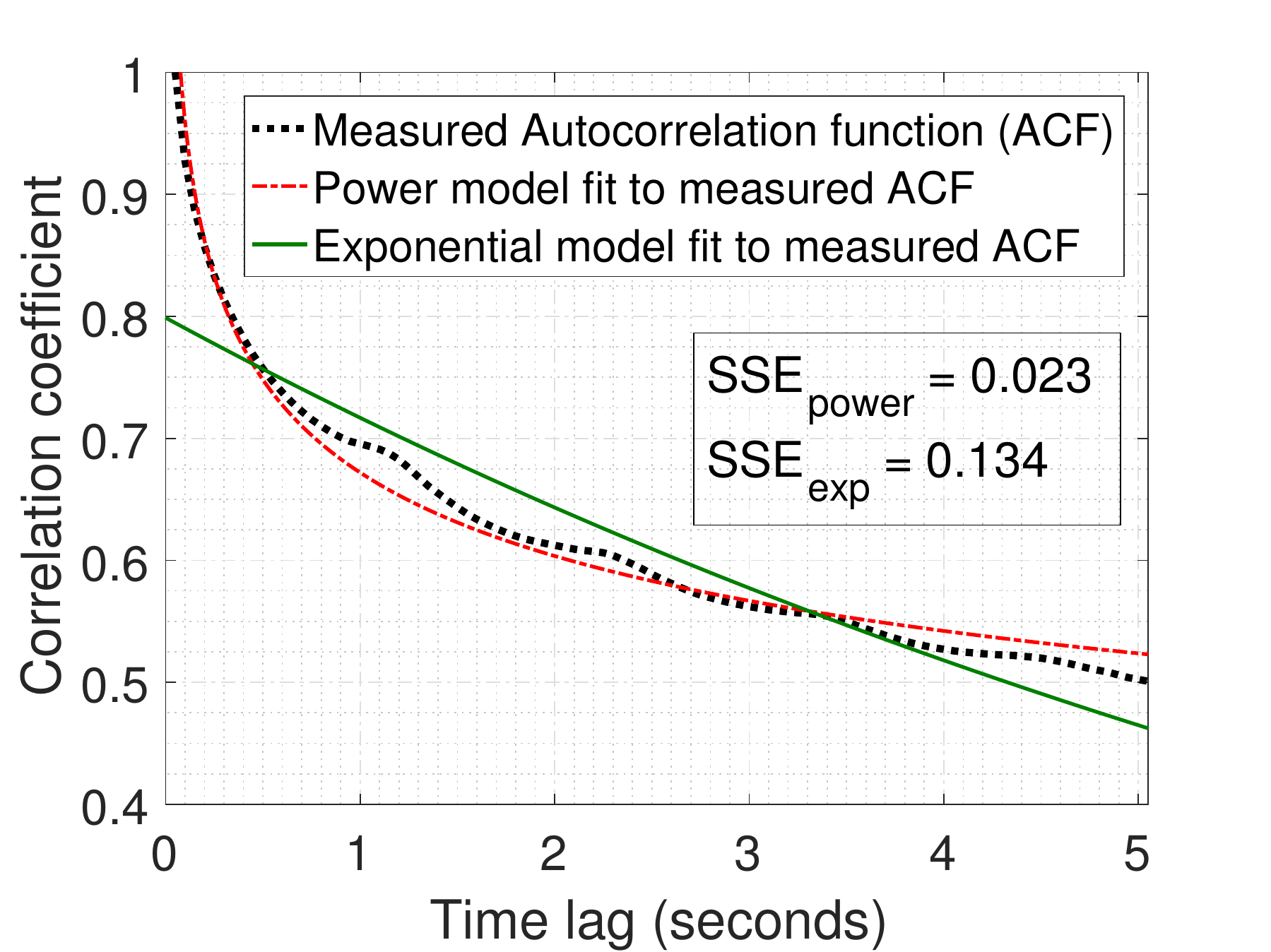}
\caption{ACF (LH--RA)}
\end{subfigure}\hfill
\begin{subfigure}{.3\textwidth}
\centering
\includegraphics[width=.8\textwidth]{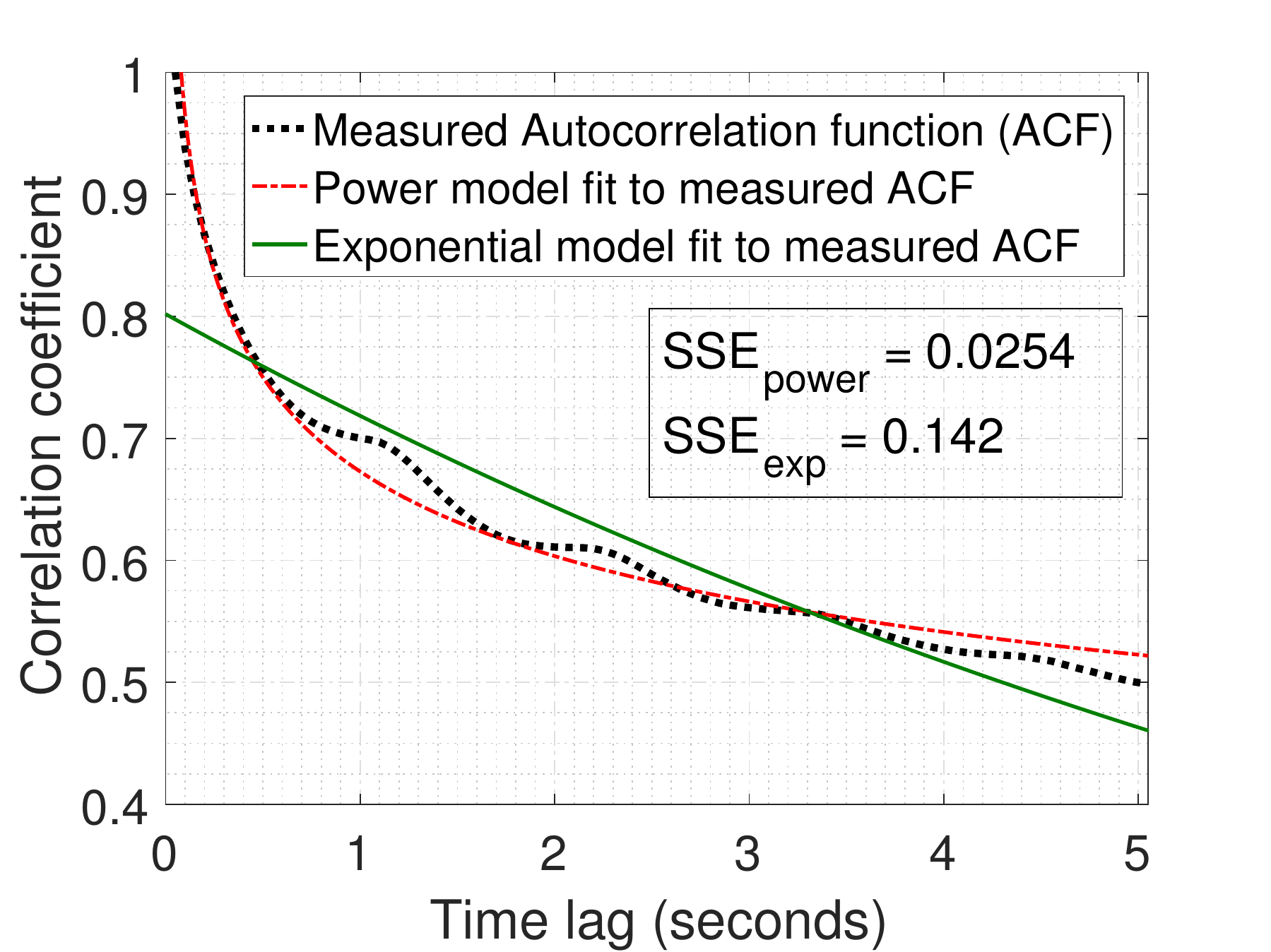}
\caption{ACF (LH--LW)}
\end{subfigure}
\caption{Power fit and exponential fit to averaged autocorrelation decay of different body-to-body channels}
\label{B2B_ACF}
\end{figure*}

\begin{figure}[!t]
\begin{subfigure}{.5\columnwidth}
\centering
\includegraphics[width=\textwidth]{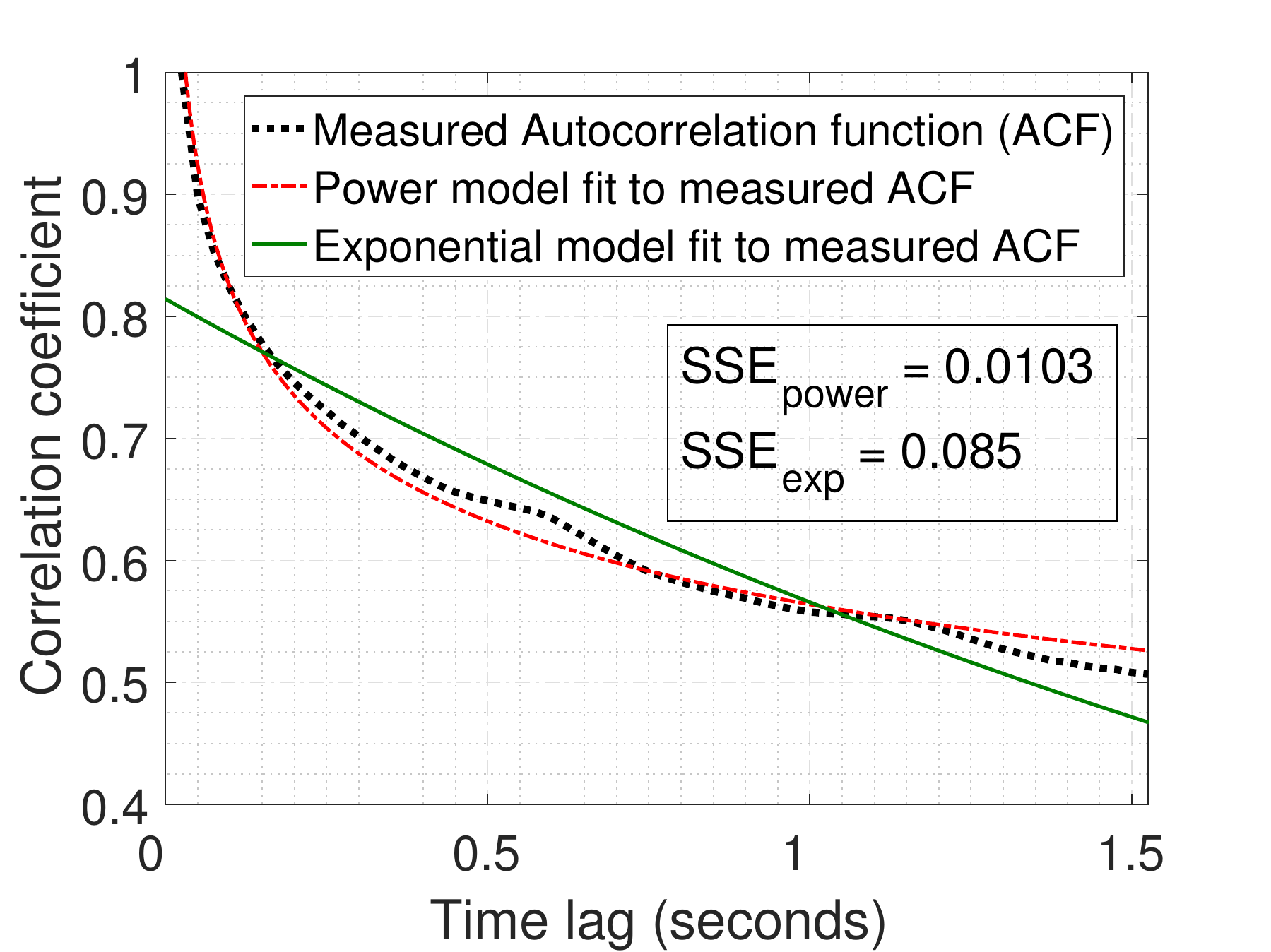}
\caption{ACF (LH--RA$_{\text{on-body}}$)}
\end{subfigure}\hfill
\begin{subfigure}{.5\columnwidth}
\centering
\includegraphics[width=\textwidth]{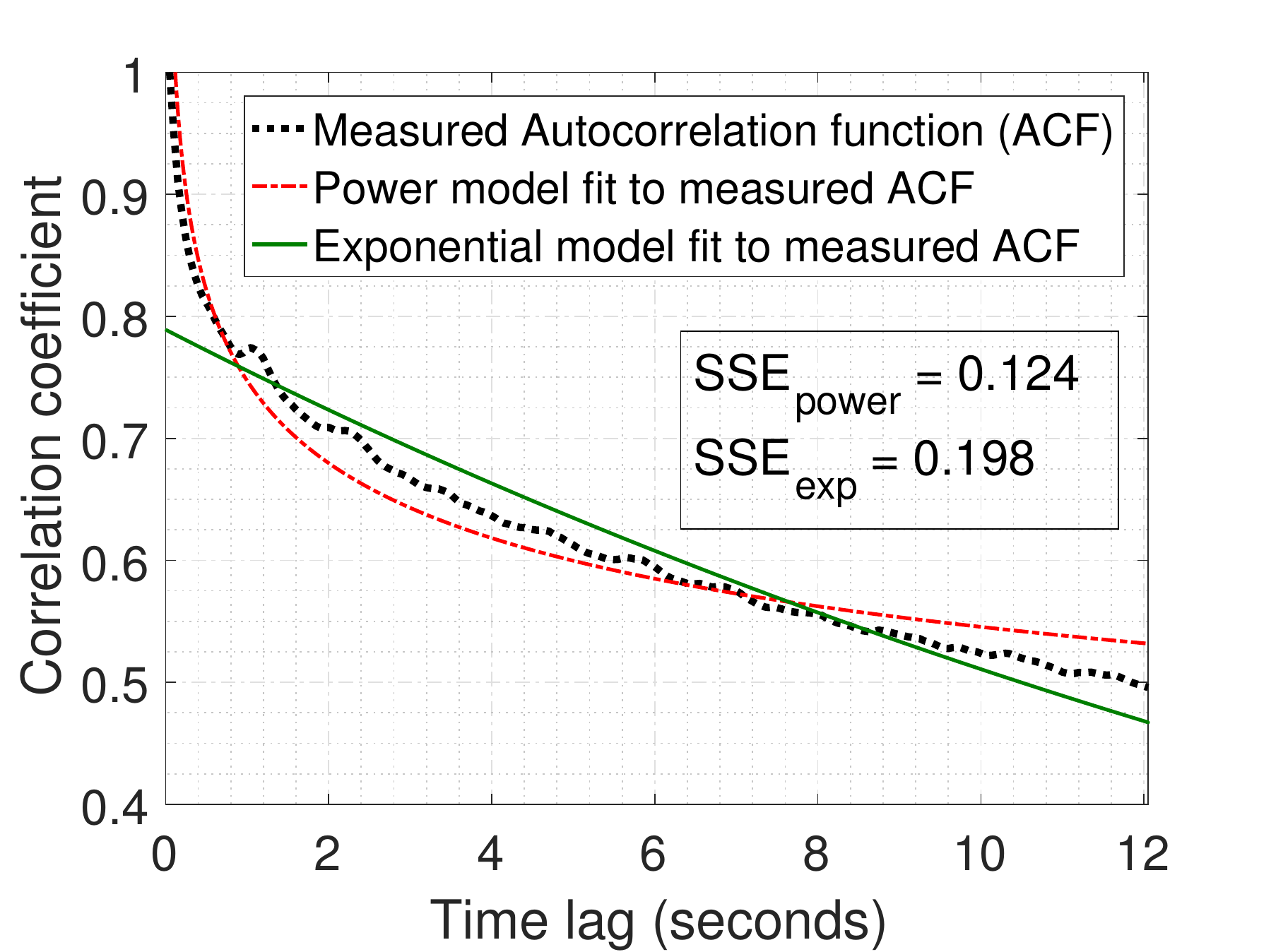}
\caption{ACF (LH--LW$_{\text{on-body}}$)}
\end{subfigure}
\caption{Power fit and exponential fit to averaged autocorrelation decay of different on-body channels}
\label{BAN_ACF}
\end{figure}
We analyze the average ACF of different BAN/BBN channels where we fit the single term exponential and power series models to the ACF decay in MATLAB, which uses the trust-region algorithm with nonlinear least-square method. 
The power and exponential fit to the measured averaged ACF for different B$2$B and on-body channels are shown in Figs. \ref{B2B_ACF} and \ref{BAN_ACF}, respectively. The models are fitted to the ACF decay till a moderate correlation coefficient of $0.5$ to measure the optimum result. We measure the goodness-of-fit with the sum of squared errors of prediction (SSE) statistic \cite{hamilton1994time}. A SSE value closer to $0$ indicates that the model has a smaller random error component, and that the fit will be more useful for prediction. 
It can be seen from Figs. \ref{B2B_ACF} and \ref{BAN_ACF} that, both the on-body and B$2$B channels show a power-like decay for the autocorrelation function (SSE closer to $0$). From that outcome, we can imply that the autocorrelation function of body-centric channels (B$2$B/on-body) has power-like decay, hence these channels possess long-range dependence. 

\subsection{Hurst exponent} 

\vspace{-1em}A more systematic approach to analyze the dependence of the channels is to estimate the Hurst exponent, which is also referred as the index of dependence. The value of Hurst exponent ($h_E$) ranges between $0$ and $1$. If $h_E=0.5$, then it indicates that there is no correlation/dependence between the points of the channel. If ($0.5<h_E<1$), then the channel characteristics is persistent (e.g., an increment/decrement is followed by another increment/decrement in the near future). And, if ($0<h_E<0.5$), then the channel characteristics is anti-persistent (e.g., an increment is followed by a decrement in the near future and vice-versa). In a nutshell, as $h_E$ moves away from $0.5$, it tends to give more information about the channel characteristics, hence the channel becomes more predictable.
\begin{figure}[!h]
\begin{subfigure}{.5\columnwidth}
\centering
\includegraphics[width=\textwidth]{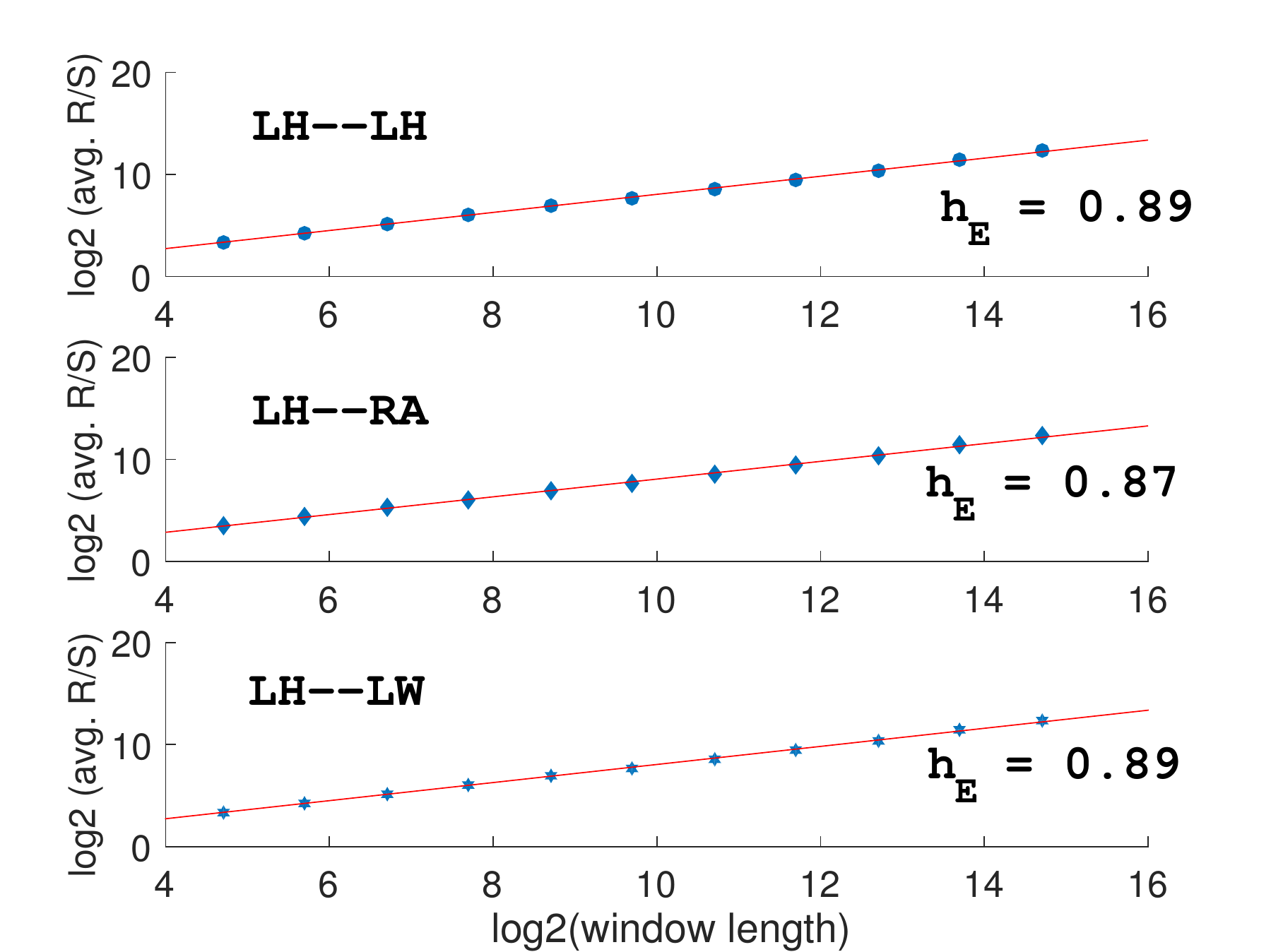}
\caption{$h_E$ (B$2$B)}
\end{subfigure}\hfill
\begin{subfigure}{.5\columnwidth}
\centering
\includegraphics[width=\textwidth]{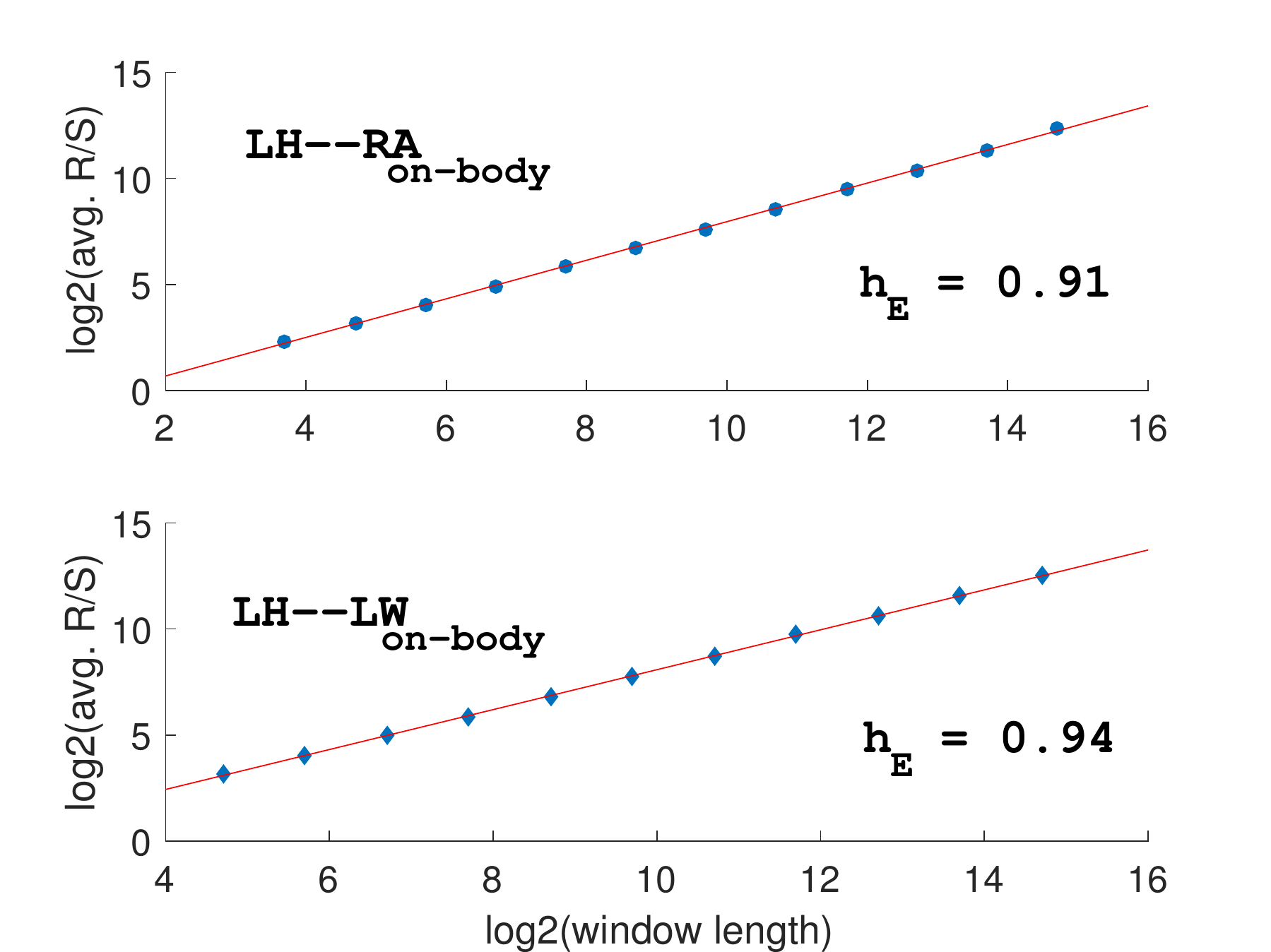}
\caption{$h_E$ (on-body)}
\end{subfigure}
\caption{Hurst regression from averaged R/S values for different B$2$B (a) and on-body (b) links with $10$ co-located BANs. Hurst exponent ($h_E$) is calculated from the slopes of the red lines.}
\label{H_e}
\end{figure}
For measuring the Hurst exponent, we follow the rescaled range (R/S) analysis method described in \cite{feder1991new}. 
We average the $E[R/S]$ value from different groups of similar B$2$B/on-body links and measure the approximate Hurst exponent for specific type of B$2$B/on-body links. The results are shown in Fig. \ref{H_e}, where all of the links are giving a higher value of Hurst index (greater than $0.5$). From these results it can be inferred that, body-centric channels (B$2$B/on-body) incorporate long-range dependence.

\vspace{-0.5em}\subsection{Long-memory and Stationarity}

\vspace{-0.5em}Beside statistical dependence, stationarity or wide-sense-stationarity (statistical properties, e.g., mean, auto-covariance, are invariant over time) is another important characteristic to estimate the predictability of a channel. From the long-memory outcome, it can be inferred that, both type of channels (on-body/B$2$B) are predictable. But we show in \cite{sshimly2018stationarity} that, B$2$B channels can possess wide-sense-stationarity (WSS) for certain period, whereas on-body channels depict non-stationary behavior. Hence, even if on-body channels can have long-memory, that memory is not useful because of the non-stationary behavior, which can produce spurious results.

\section{Conclusion}

\vspace{-0.5em}We show that, body-centric channels (on-body/B$2$B) can possess long-memory. However, only B$2$B links can be utilized for reliable predictive analysis due to their WSS property. 


%

\vspace{-0.5em}
\bibliographystyle{IEEEtran}
\bibliography{Reference}

\begin{thebibliography}{1}
\providecommand{\url}[1]{#1}
\csname url@samestyle\endcsname
\providecommand{\newblock}{\relax}
\providecommand{\bibinfo}[2]{#2}
\providecommand{\BIBentrySTDinterwordspacing}{\spaceskip=0pt\relax}
\providecommand{\BIBentryALTinterwordstretchfactor}{4}
\providecommand{\BIBentryALTinterwordspacing}{\spaceskip=\fontdimen2\font plus
\BIBentryALTinterwordstretchfactor\fontdimen3\font minus
  \fontdimen4\font\relax}
\providecommand{\BIBforeignlanguage}[2]{{%
\expandafter\ifx\csname l@#1\endcsname\relax
\typeout{** WARNING: IEEEtran.bst: No hyphenation pattern has been}%
\typeout{** loaded for the language `#1'. Using the pattern for}%
\typeout{** the default language instead.}%
\else
\language=\csname l@#1\endcsname
\fi
#2}}
\providecommand{\BIBdecl}{\relax}
\BIBdecl

\bibitem{sshimly2018}
S.~M. Shimly, D.~B. Smith, and S.~Movassaghi, ``Experimentally-based
  cross-layer optimization across multiple wireless body area networks,''
  \emph{arXivpreprint arXiv: 000.0000}, 2018.

\bibitem{hanlen2010open}
L.~Hanlen, V.~Chaganti, B.~Gilbert, D.~Rodda, T.~Lamahewa, and D.~Smith,
  ``Open-source testbed for body area networks: 200 sample/sec, 12 hrs
  continuous measurement,'' in \emph{IEEE 21st International Symposium on
  Personal, Indoor and Mobile Radio Communications Workshops (PIMRC Workshops),
  Turkey}, Sep, 2010, pp. 66--71.

\bibitem{smith2012body}
D.~Smith, L.~Hanlen, D.~Rodda, B.~Gilbert, J.~Dong, and V.~Chaganti, ``Body
  area network radio channel measurement set,'' \emph{URL:
  http://doi.org/10.4225/08/5947409d34552}, 2012.

\bibitem{hamilton1994time}
J.~D. Hamilton, \emph{Time series analysis}.\hskip 1em plus 0.5em minus
  0.4em\relax Princeton university press Princeton, 1994, vol.~2.

\bibitem{feder1991new}
J.~Feder and P.~P. Fractals, ``New york, 1988,'' \emph{Google Scholar}, 1991.

\bibitem{sshimly2018stationarity}
S.~M. Shimly, D.~B. Smith, and S.~Movassaghi, ``Wide-sense-stationarity of
  everyday wireless channels for body-to-body networks,'' in \emph{the IEEE
  International Conference on Communications (accepted), available on arxiv},
  2018.

\end{thebibliography}

\vspace{-1em}

\end{document}